\documentclass[12pt]{iopart}
\usepackage{subfigure}
\usepackage{amssymb}
\usepackage{amsfonts}
\usepackage{graphicx}
\usepackage[usenames,dvipsnames]{color}
\usepackage[all]{xy}
\usepackage[normalem]{ulem}
\usepackage{scrextend}

\begin{document}
\newcommand{\av}[1]{{\color{blue}$\clubsuit$#1}}
\newcommand{\cor}[1]{{\color{red}$\spadesuit$#1}}
\newcommand{\pb}[1]{{\color{orange}$\spadesuit$#1}}
\newcommand{\jb}[1]{{\color{RoyalPurple}$\diamondsuit$#1}}
\newcommand{\erf}[1]{{\color{green}$\heartsuit$#1}}

\title[Energy ordering of grain boundaries in Cr$_2$O$_3$]{Energy ordering of grain boundaries in Cr$_2$O$_3$: Insights from theory}
\author{A.G. Van Der Geest$^{1,2}$, M. M. Islam$^{1,3}$, T. Couvant$^4$,  and B. Diawara$^1$}
\date{\today}
\address{$^1$Laboratoire de Physico-Chimie des Surfaces, CNRS-ENSCP (UMR7045), Ecole Nationale Sup\'erieure de Chimie de Paris, F-75005 Paris, France}
\address{$^2$Department of Physics, Applied Physics, and Astronomy, University of Binghamton, State University of New York, PO Box 6000, Binghamton, New York 13902-6000, USA}
\address{$^3$Mulliken Center for Theoretical Chemistry, Institut f\"ur Physikalische und Theoretische Chemie, Universit\"at Bonn, Beringstr. 4, 53115 Bonn, Germany }
\address{$^4$ EDF R\&D, Centre des Renardi\`ers, Route de Sens 77818, Moret-sur-Loing, France}

\begin{abstract}
\noindent{The grain boundaries, GBs, of corundum Cr$_2$O$_3$ are known to play an important role in the diffusion of ions within the oxide, which is an important phenomenon for the corrosion of the stainless steels.
The extent of the growth of oxide layers in stainless steel depends upon which interfaces are preferred within Cr$_2$O$_3$.  Therefore, we have constructed four different grain boundary planes (rhombohedral, basal, prismatic and pyramidal) and their various associated interface symmetries known in literature for corundum Al$_2$O$_3$. Their structural, electronic, and energetic properties are investigated theoretically with periodic calculations using the DFT+U approach. We find that the prismatic screw GB with a Cr-O plane interface is the energetically preferred GB with the rhombohedral GB with screw symmetry and Cr vacancy termination being the second energetically preferred GB.  The increase of the number of in-plane Cr atoms at the interface of prismatic GB enhances the stability which is also evident in the electronic density of states.}


\end{abstract}

\pacs{61.72.Mm}

\maketitle

\section{Introduction}

In many applications, the ability of the Cr$_2$O$_3$ thin films to prevent corrosion is used to improve the corrosion resistance of metallic structures. This includes the inclusion of Cr into stainless-steels and other metal alloys with the intent of the Cr forming Cr$_2$O$_3$ scale at interfaces with oxygen or water. In particular austenitic alloys (austenitic stainless steels and nickel-based alloys) are used in the nuclear industry as components in pressurized water reactors (PWR). On these alloys, the native Cr$_2$O$_3$ films inhibit the diffusion of ions of the underlying material and oxygen from the environment and thus preventing oxidation. 

Understanding the real mechanisms involved in the growth of protective oxide layer is an important issue. Based on the identified mechanisms, modeling, and computer simulation of the growth of oxide layers allow for the long term prediction of the mechanical and chemical behavior of metallic materials in corrosive environments. Among the various approaches, atomistic modeling is particularly interesting, providing access to the effect of the local chemistry and structure of the film, while also allowing for a fine-tuning of the interplay of the elementary processes involved during oxidation~\cite{Diawara2010}.  Such atomistic models need local values of the diffusion coefficient to account for the local chemical and topological environments. Some attempts have been made to calculate  the diffusion coefficients by \emph{ab initio} calculations of both Cr$_2$O$_3$ and at the interfaces in a complex Cr/Cr$_2$O$_3$/Cr(OH)$_3$ system~\cite{Yu2012}. In all these works the oxide scale is considered as uniform, while numerous studies have shown the presence of grain boundaries (GBs) in the oxide affects ion diffusion in the oxide. As is shown by Tsai \emph{et al.} \cite{Tsai1996}, Cr and O diffusing through GBs have different diffusion rates compared to bulk Cr$_2$O$_3$. Therefore to appropriately study diffusion in Cr$_2$O$_3$ GBs must be considered.


It is important to understand the structure of the GB interfaces that exist within the oxide to study diffusion at GBs using \emph{ab initio} models.  However, there are relatively few studies of these structures for Cr$_2$O$_3$.   Fang \emph{et al.}\cite{Fang2012} have used \emph{ab initio} techniques to study temperature dependent energies of a prismatic interface and two separate Basal interfaces along with the segregation energies of several dopants at these interfaces.
Catlow \emph{et al.}\cite{Catlow1989} used empirical potentials to study several twin boundary conditions.  While Cr$_2$O$_3$ grain boundary structures have not been heavily discussed within \emph{ab initio} literature, alumina, which also possesses the corundum structure, has been intensively studied.  
Specifically, several grain boundary orientations within Al$_2$O$_3$ have been determined within a density functional theory, DFT, approach combined with experimental results. 
Marinopoulos \emph{et al.} have studied the Rhombohedral~\cite{Marinopoulos2000} and Basal~\cite{Marinopoulos2001} plane GBs with their common symmetries, while Fabris \emph{et al.} have studied the Prismatic~\cite{Fabris2002} and Pyramidal~\cite{Fabris2001} plane and symmetries. 

In the present study, we have considered various possible twin interfacial structures at Cr$_2$O$_3$ GBs using 4 GB planes with several associated symmetries for a total of 10 interface systems as determined in Al$_2$O$_3$ literature.   We have calculated the structural, energetic and electronic properties of Cr$_2$O$_3$ GBs  and evaluated the relative stabilty for the first time.

\section{Computational Method}

The calculations to investigate the atomic structures and system energies of the grain boundaries were performed using DFT+U as implemented within the \texttt{VASP}\cite{Kresse1996b,Kresse1996} software package.   A spin polarized, GGA-PW91 functional~\cite{perdew1993}, within an augmented plane wave framework~\cite{Blochl1994} was used with an energy cutoff of 520 eV as optimized for the bulk properties of Cr$_2$O$_3$ in the present study.
Monkhorst-Pack\cite{Monkhorst1976} k point grid of 4$\times$4$\times$4 for the bulk Cr$_2$O$_3$ and 4$\times$4$\times$1 for the grain boundaries were used. For Cr, a +U correction was applied for the strongly localized correlated $d$ electrons based upon 
the approach of Dudarev \emph{et al.} \cite{dudarev1998} with an effective on-site coulomb interaction parameter of 5 eV as defined in \cite{Rohrbach2004}. Geometry optimizations were performed with the conjugate 
gradient algorithm within an energy difference of $10^{-4}$ eV.

Following the DFT calculation the interfacial energies for a specific grain boundary were calculated by
\begin{equation}\label{eqn:eint}
E_{int}=\frac{(E_{GB}-n E_{bulk})}{2A}.
\end{equation}
Here E$_{GB}$ is the total energy of the interface, E$_{bulk}$ is the energy of a single formula unit of Cr$_2$O$_3$ in the bulk, $n$ is the number of formula units in the GB, and A is the area of interface plane in the supercell.  Effectively, this compares the GB to an infinitely large bulk system that is the most stable isomorph of Cr$_2$O$_3$.   This reduces the inherent difficulty of comparing energies between different interfaces by comparing their difference from the bulk state.  The lower the interface energy the more energetically stable the GB is. By this way, the energetic stability between the various GB can be compared as was done for Al$_2$O$_3$ systems~\cite{Marinopoulos2000, Marinopoulos2001, Fabris2002, Fabris2001} and prismatic and basal interfaces in Cr$_2$O$_3$~\cite{Fang2012}.

\section{Construction of Grain Boundary Models}
Using the planes and symmetries determined within Al$_2$O$_3$ it is possible to sample realistic planes and symmetries for Cr$_2$O$_3$.  As such these structures were constructed for Cr$_2$O$_3$.  It is worth noting that all of the grain boundaries studied are of the form that the two grains have the same interfacial planes represented as (abcd)$\left|\right|$(abcd) where (abcd) is the interface plane of a given grain.  

\begin{figure}
\centering 
\includegraphics[width=0.4\textwidth]{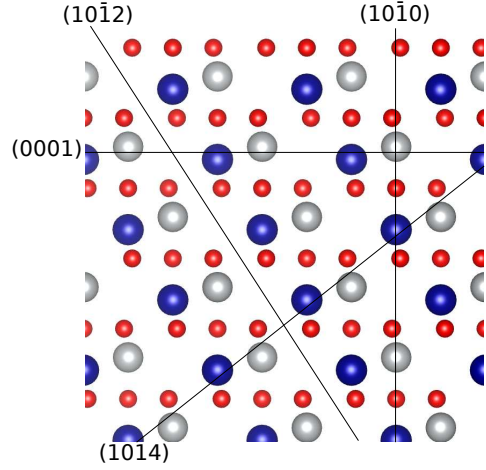}
\caption{The direction of the grain boundary planes considered in this paper in relation to the hexagonal conventional bulk cell of Cr$_2$O$_3$.  Here red represents O, blue spin up Cr, and white is spin down Cr (Color Online).}\label{img:planes}
\end{figure}

\begin{table*}[hbtp]
    \caption{Each grain boundary type considered shown with  its interface plane and the miller indices of the vectors used to construct the cell. The symmetries for each grain boundary type are then given along with the relative shifts in each direction and with the interfacial energy $E_{int}$. 
The parenthesis represents the surface termination of the symmetry. V stands for Cr vacancy termination, O stands for O termination, Cr is Cr termination, and a lack of parenthesis is a Cr-O plane. }
{\centering
 \begin{tabular}{@{}l c c c c c c } 
      \br
      Grain Boundary   &Label &$\vec{e_1}$&$\vec{e_2}$&$\vec{e_3}$&$E_{int}$ (Cr$_2$O$_3$) &$E_{int}$ (Al$_2$O$_3$)\\
       \qquad Symmetry && \qquad T$_1$ &\qquad T$_2$ &\qquad  T$_3$ &(J/m$^2$)&(J/m$^2$)\\
     \mr
       Rhom.   (1012)     & & [10$\bar{1}$1] &  [$\bar{1}$2$\bar{1}$0]  &[$\bar{5}$052]  &&\\
         \qquad Glide(V) & rG(V)&\qquad0.0&\qquad0.5&\qquad 0.0& 1.93& 3.34\cite{Marinopoulos2000} \\
         \qquad Glide(O)  &rG(O)&\qquad0.0&\qquad0.5&\qquad 0.125$^a$&1.08&1.35\cite{Marinopoulos2000}\\
         \qquad Screw(V) &rS(V)&\qquad0.5&\qquad0.0&\qquad 0.0&0.33&0.63\cite{Marinopoulos2000}\\
       Basal (0001)                        & & [$\bar{1}$010]  & [1$\bar{2}$10]  & [0001] &&\\
         \qquad Rotational (Cr)& bR(Cr)&\qquad 0.5  &\qquad 0.0&\qquad 0.0& 0.61&0.73\cite{Marinopoulos2001}\\
         \qquad Mirror (O)   &    bM(O)&\qquad0.0   &\qquad0.0 &\qquad 0.0&  1.35&1.99\cite{Marinopoulos2001}\\
         \qquad Glide-Mirror (O)$^b$& bG(O)&\qquad0.0&\qquad0.$\bar{1}$& \qquad0.0&1.59&2.63\cite{Marinopoulos2001}\\
      Prismatic  (10$\bar{1}$0)                &  & [$\bar{1}$2$\bar{1}$0]  & [0001]  & [1010] &&\\
         \qquad Glide       &  prG&\qquad0.5   &\qquad0.0                & \qquad 0.0&0.40&0.49\cite{Fabris2002}\\
         \qquad Screw      &  prS&\qquad0.5  &\qquad0.3$\bar{3}$&\qquad 0.0&0.23&0.30\cite{Fabris2002}\\
       Pyramidal  (10$\bar{1}$4)            &  & [20$\bar{2}$$\bar{1}$]  & [1$\bar{2}$10]  &[50$\bar{5}$4] &&\\
         \qquad Glide(Cr) & pyG(Cr)&\qquad 0.25&\qquad 0.5&\qquad 0.0&1.14&1.88\cite{Fabris2001}\\
         \qquad Glide(O)  & pyG(O)&\qquad 0.5   &\qquad 0.5&\qquad 0.07$^a$&1.14&2.44\cite{Fabris2001}\\
      \br
   \end{tabular}
   }
   \begin{indented}
   \item[]$^a$Offset after optimization of the $\vec{e_3}$ direction.
   \item[]$^b$"Bulk" regions of the system are constrained from shifting.
   \end{indented}
   \label{tab:gbenergy2}
\end{table*}

For each of the systems constructed we took the original bulk corundum structure and created a mirror grain across a GB interface plane (a list of GBs are  in table \ref{tab:gbenergy2} and shown in Figure  \ref{img:planes}) then applied shifts to the new mirrored grain parallel to the interface plane to create the symmetry groups.  Here this was done within supercells with lattice vectors $\vec{e_1}$, $\vec{e_2}$, and $\vec{e_3}$ as defined for each of the rhombohedral, basal, prismatic and pyramidal interfaces in table \ref{tab:gbenergy2}.  The vectors $\vec{e_1}$ and $\vec{e_2}$ are parallel to the interface plane, while $\vec{e_3}$ is perpendicular to the interface.  The in-plane shifts were applied by pre-factors T$_1$ and T$_2$ (also in table \ref{tab:gbenergy2}) in the $\vec{e_1}$ and $\vec{e_2}$ directions, while T$_3$ was applied for alternately terminated interfaces.  The mirror symmetry is obtained when $T_1=T_2=0$, while the vector pre-factors for the rotational, glide, and screw symmetries symmetry vary with surface direction.  

\begin{figure}
\centering 
\includegraphics[width=0.45 \textwidth]{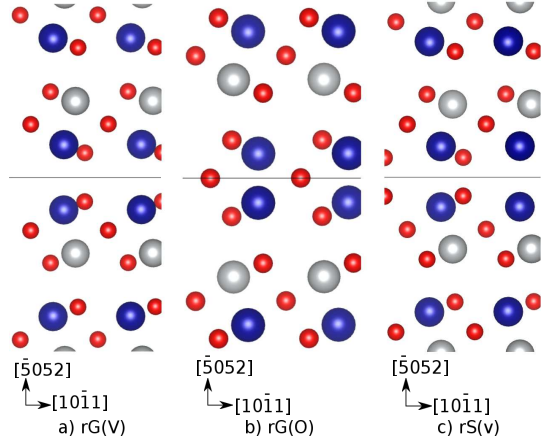}
\caption{The three rhombohedral grain boundary planes: a) Glide (G) with vacancy termination b) Glide (G) with oxygen termination, and c) Screw (S) with vacancy termination. Here red represents O, blue spin up Cr, and white is spin down Cr. All three GBs are viewed along the [$\bar{1}$2$\bar{1}$0] direction (Color Online).  }\label{img:rhom}
\end{figure}

\begin{figure}
\centering 
\includegraphics[width=0.425\textwidth]{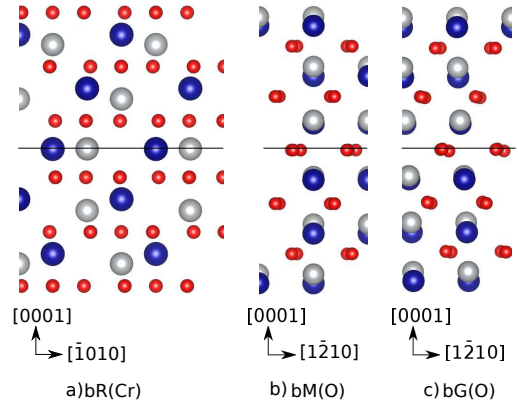}
\caption{The three basal grain boundary planes: a) Rotational (R) symmetry with Cr termination, b) Mirror (M) symmetry with O termination, and c) Glide (G) symmetry with O termination. Here red represents O, blue spin up Cr, and white is spin down Cr. The  rotational GBs is viewed along the [1$\bar{2}$10] direction, while the glide and mirror GBs are viewed from the [$\bar{1}$010] direction to show the shift in the vertical arrangement of Cr atoms between the two GBs as described in the text (Color Online).  }\label{img:basal}
\end{figure}

\begin{figure}
\centering 
\includegraphics[width=0.475\textwidth]{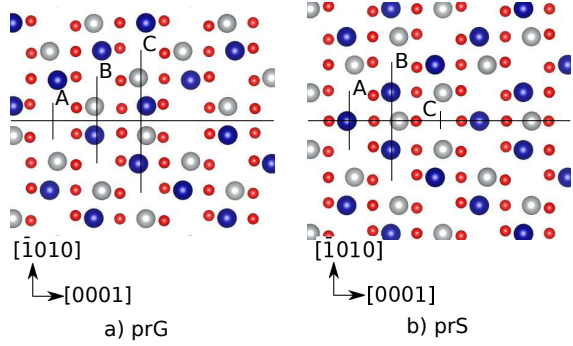}
\caption{The two prismatic grain boundary planes: a) Screw (S) symmetry and b) Glide (G) symmetry. Here red represents O, blue spin up Cr, and white is spin down Cr.  All three GBs are viewed along the [$\bar{1}$2$\bar{1}$0] direction (Color Online).}\label{img:pris}
\end{figure}

\begin{figure}
\centering 
\includegraphics[width=0.3\textwidth]{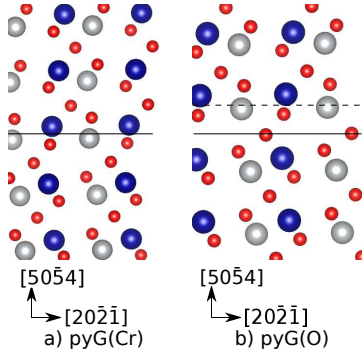}
\caption{The two pyramidal grain boundary planes: a) Glide (G) symmetry with Cr termination symmetry and b) Glide (G) symmetry with O termination. Here red represents O, green represents spin up Cr, and blue is spin down Cr (Color Online). }\label{img:pyra}
\end{figure}

Every rectangular cell is constructed to contain two interfaces due to the periodic symmetry of the cells.  It is worth noting that both the [$\bar{5}$052] and the [50$\bar{5}$4] directions for the Rhombohedral and Pyramidal directions respectively require more than 20 atomic planes to be periodic in the bulk system.  However, the cell size can be reduced due to the two sides of the grain boundary having the same angle between $\vec{e_3}$ and a bulk Cr$_2$O$_3$ lattice vector. All the GB interfaces are illustrated in Figures \ref{img:rhom} (rhombohedral), \ref{img:basal} (basal), \ref{img:pris} (prismatic) and \ref{img:pyra}(pyramidal). 

The major distinction between the Al$_2$O$_3$ GBs and the Cr$_2$O$_3$ GBs is the consideration of spin at the interface for Cr$_2$O$_3$.  Alumina is non-magnetic while chromia is antiferromagnetic.  The most favorable spin orientation in the bulk that accounts for the chromia antiferromagnetism is with alternating spins within each Cr layer (defined parallel to the (0001) plane) with the atom with a lower $\vec{z}$ coordinate being  spin up and the atom with a higher $\vec{z}$ coordinate being spin down~\cite{Rohrbach2004} as shown in Figure \ref{img:planes}.  The combination of grain boundaries of different planes through this bulk spin structure can make predicting the preferred spin state difficult.  Therefore, several spin states were tried for each interface with the most energetically preferred spin state used for grain boundary.

All the spin states used can be arranged into three groups.  The first group is where the spin structures are aligned as if they were within a continuous bulk structure.  Two examples of this are seen for  the prismatic GB with screw symmetry in Figure \ref{img:pris}b and the Cr terminated pyramidal GB with glide symmetry shown in Figure \ref{img:pyra}a.  However, due to the symmetry of the interfaces, this is often not a reasonable spin configuration.    This leads to the second type of spin state where the spins at the interface are the inverse of spins expected in the next row of a single grain.  This will be referred to as the two grains being anti-aligned with each other.  The rhombohedral GB with vacancy terminated screw symmetry is an example of anti-aligned spin states as seen in Figure \ref{img:rhom}c.   The final spin state is shown in Figure \ref{img:basal}a, where the  basal GB with rotational symmetry has a horizontally shifted spin state between the two grains that is neither aligned nor anti-aligned.  It can be seen that the spin orientation for the basal GB has been shifted by one Cr atom in the [$\bar{1}$010] direction in relation to what would exist for a spin aligned system like that seen for the prismatic screw GB.

\section{Results and Discussion}
To aid in the understand of the energy ordering among the various GB planes we first discuss the structural details, the changes in the electronic density of states, DOS, of all the GB interfaces of Cr$_2$O$_3$, and compare the energetic stability among them.


\begin{figure*}
\centering
\subfigure[rG(V)]{\label{img:DOSrglidev}\includegraphics[width=0.31\textwidth]{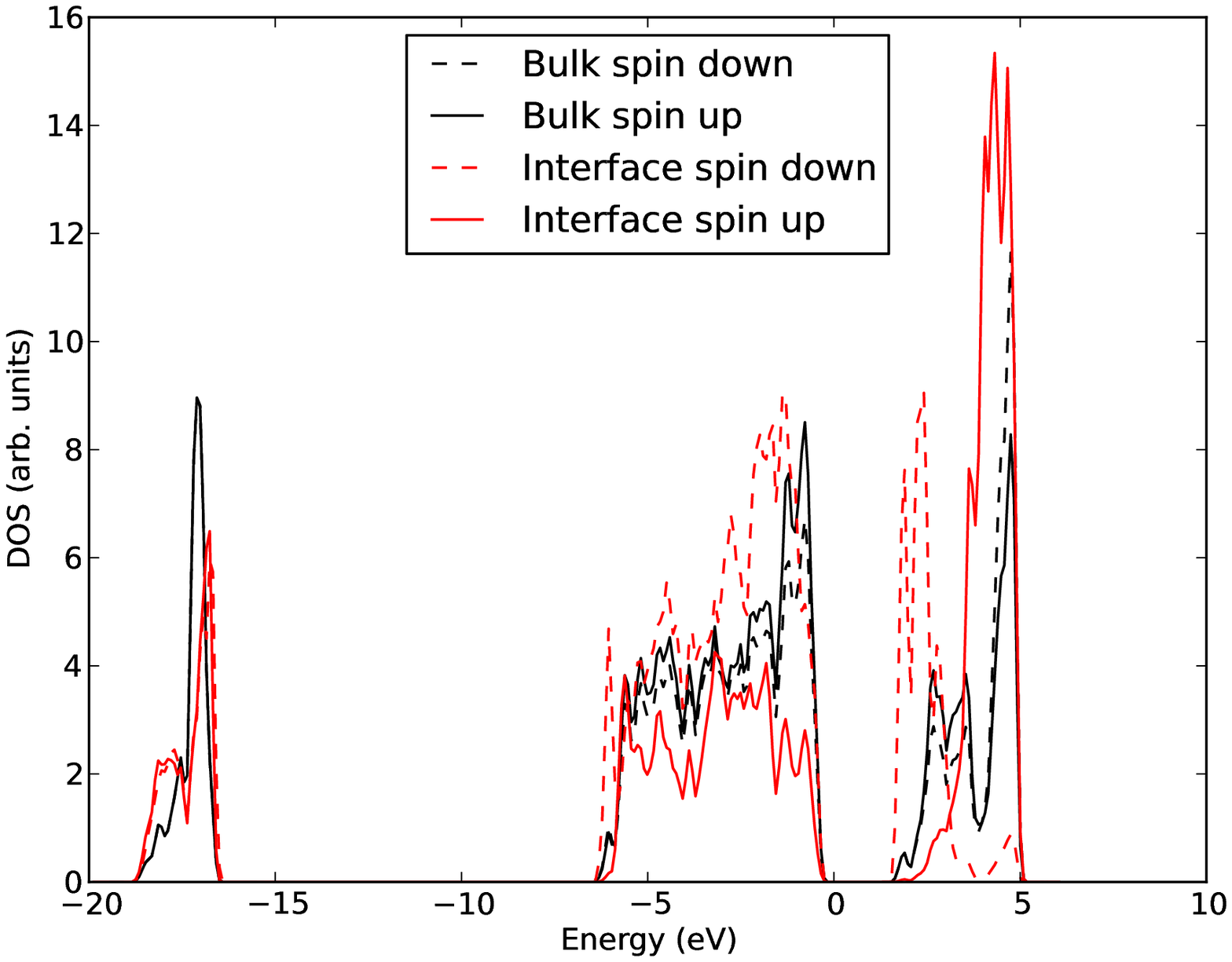}}
\subfigure[rG(O)]{\label{img:DOSrglideo}\includegraphics[width=0.31\textwidth]{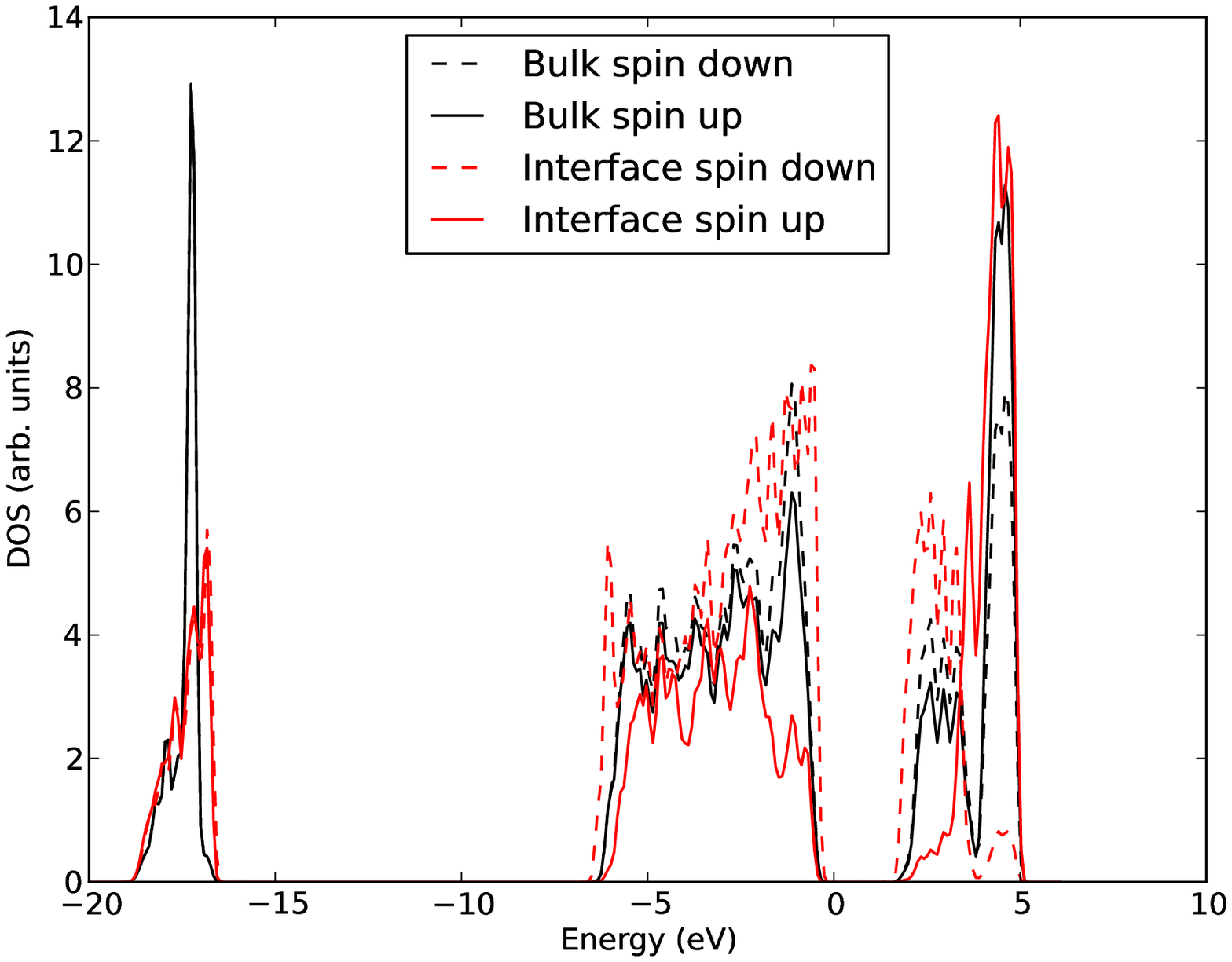}}
\subfigure[rS(V)]{\label{img:DOSrscrewv}\includegraphics[width=0.31\textwidth]{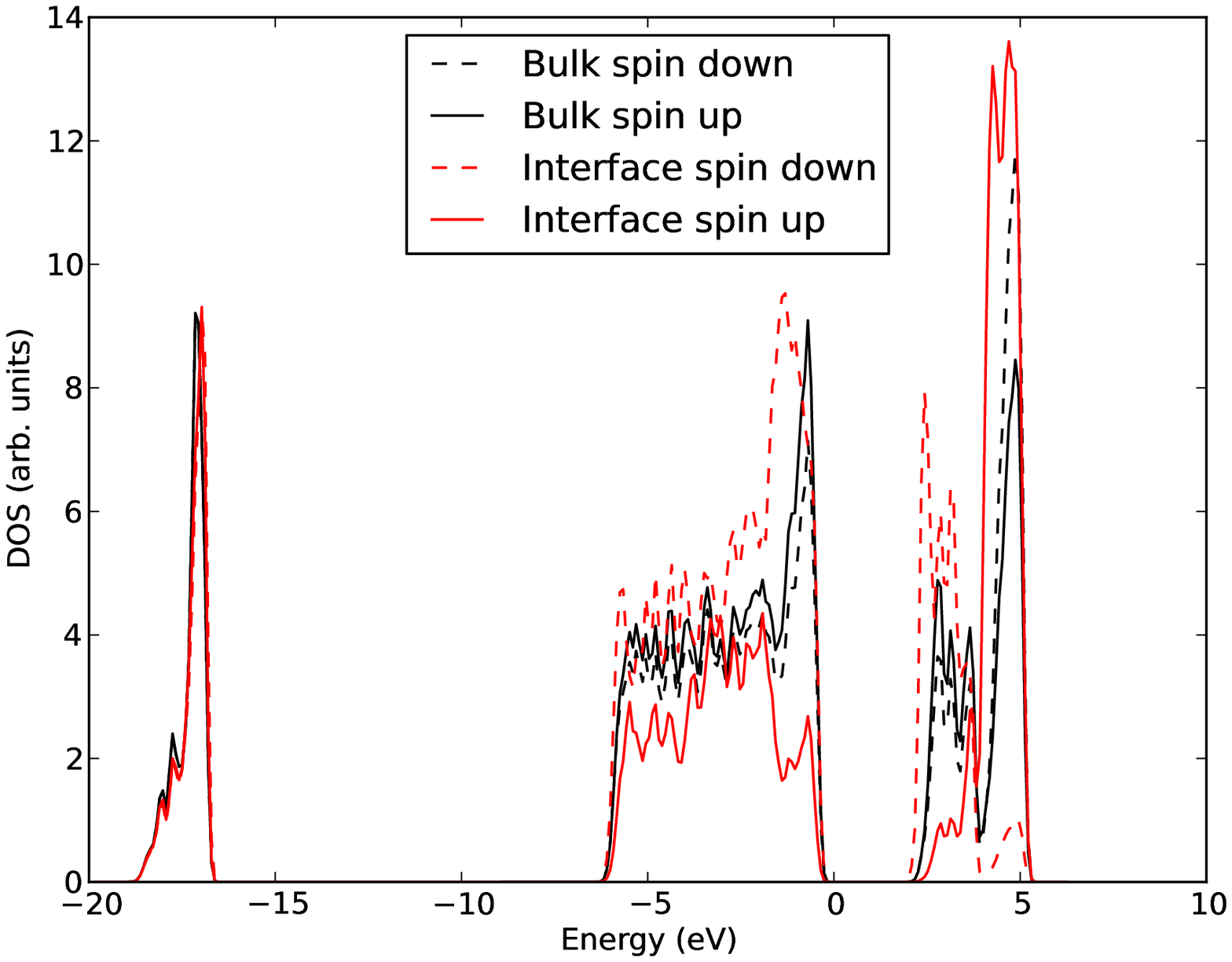}}
\subfigure[bM(O)]{\label{img:DOSbglide}\includegraphics[width=0.31\textwidth]{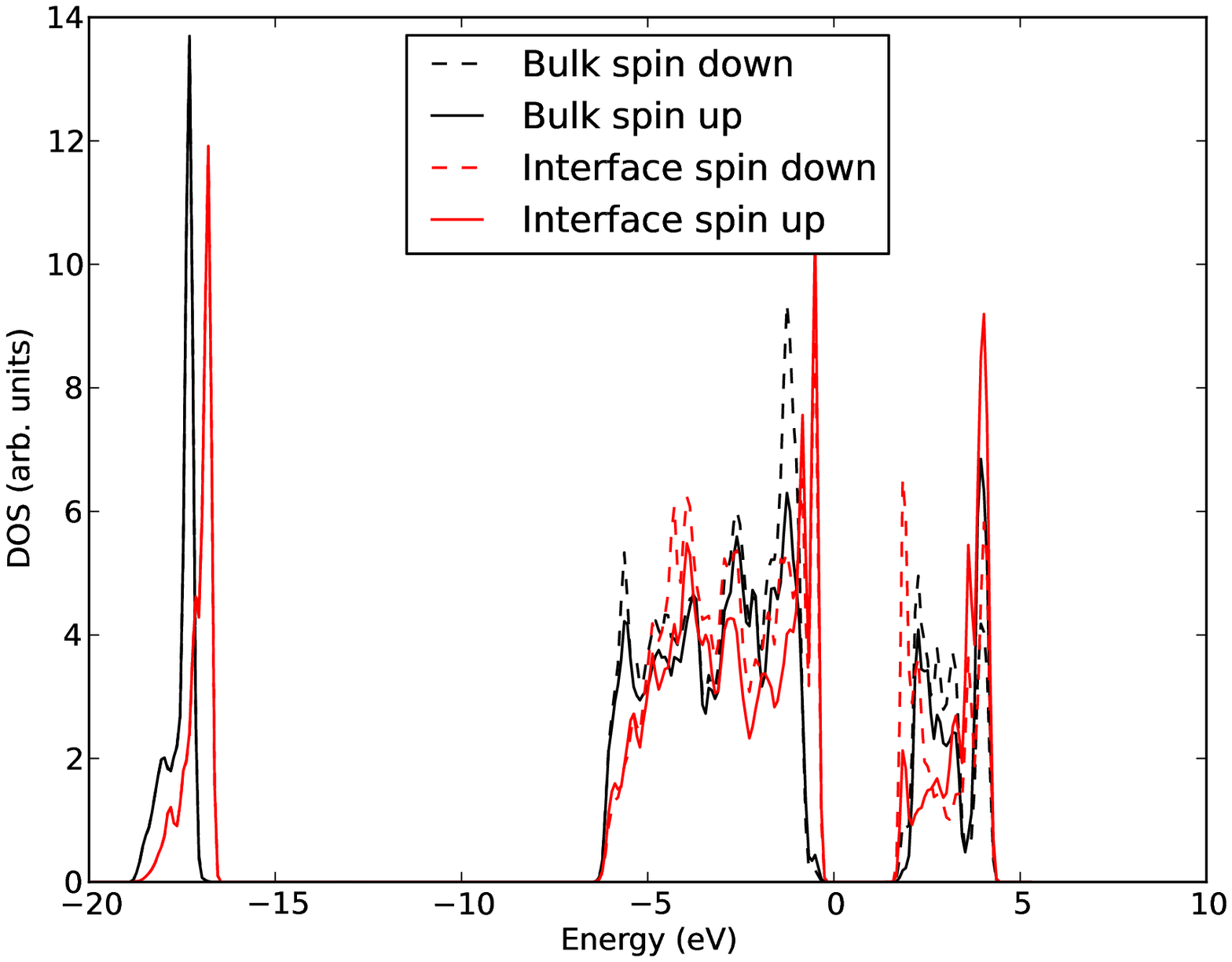}}
\subfigure[bG(O)]{\label{img:DOSbmirror}\includegraphics[width=0.31\textwidth]{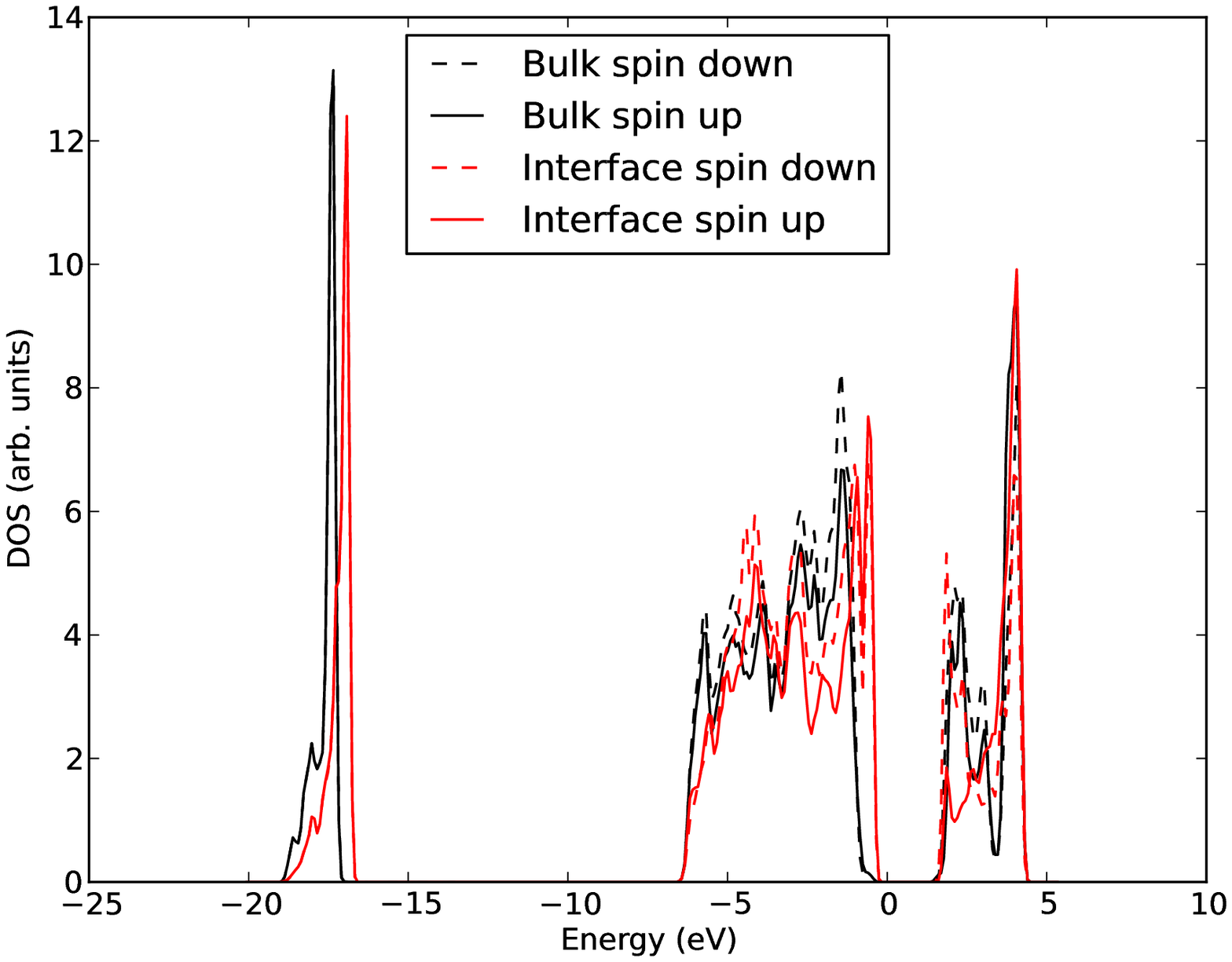}}
\subfigure[bR(Cr)]{\label{img:DOSbrot}\includegraphics[width=0.31\textwidth]{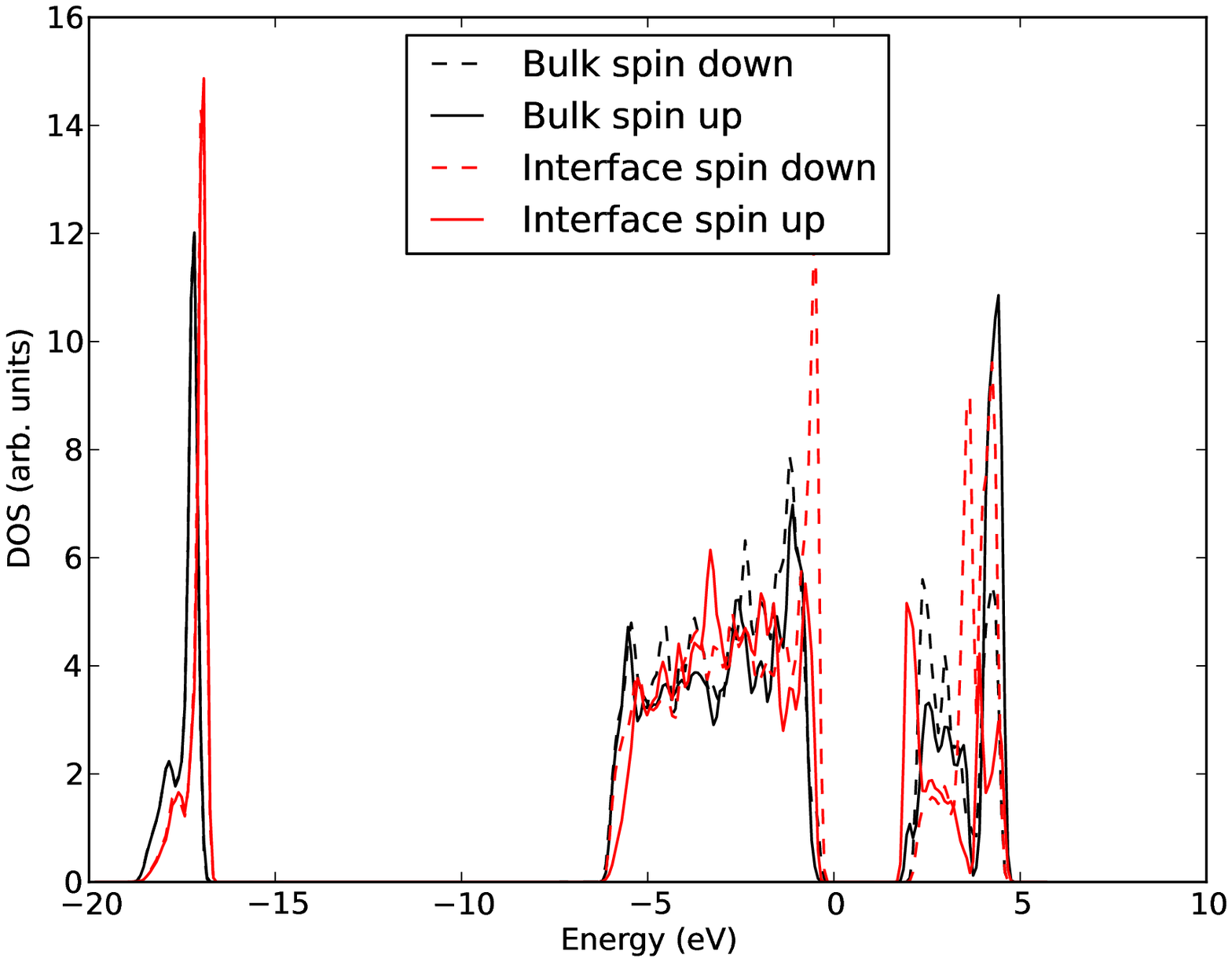}}
\subfigure[prS]{\label{img:DOSprisS}\includegraphics[width=0.31\textwidth]{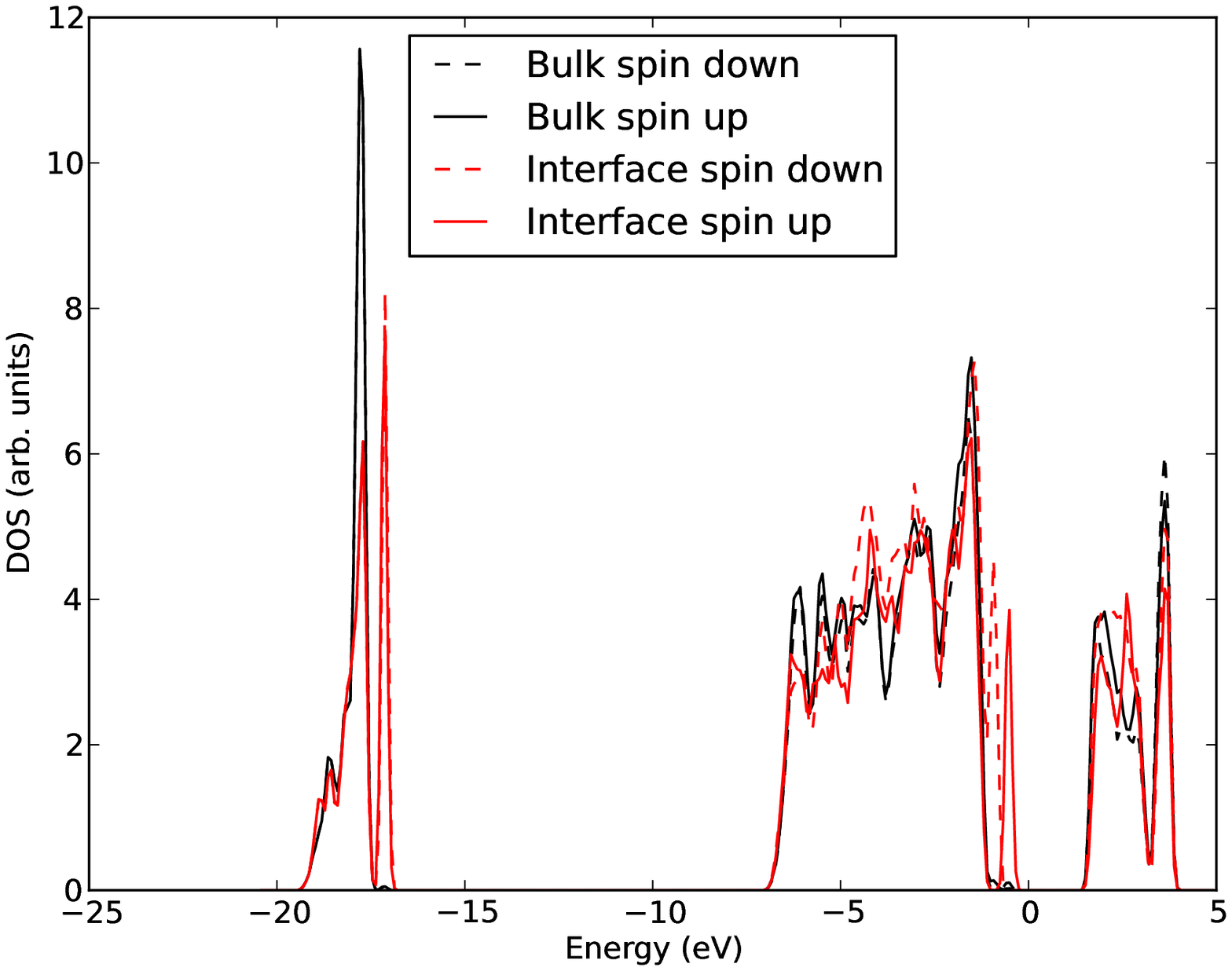}}
\subfigure[prG]{\label{img:DOSprisG}\includegraphics[width=0.31\textwidth]{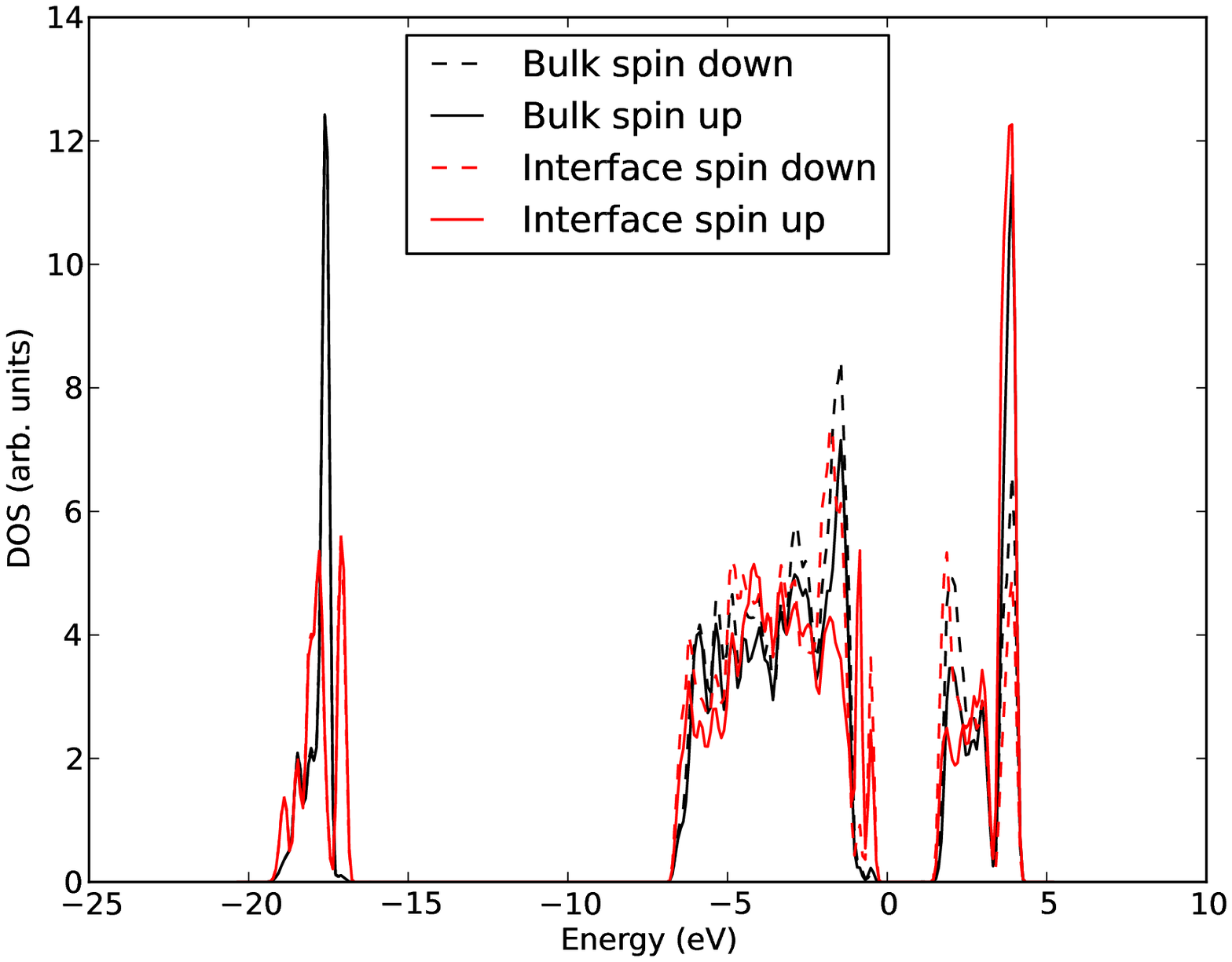}}
\subfigure[pyG(Cr)]{\label{img:DOSpyraS}\includegraphics[width=0.31\textwidth]{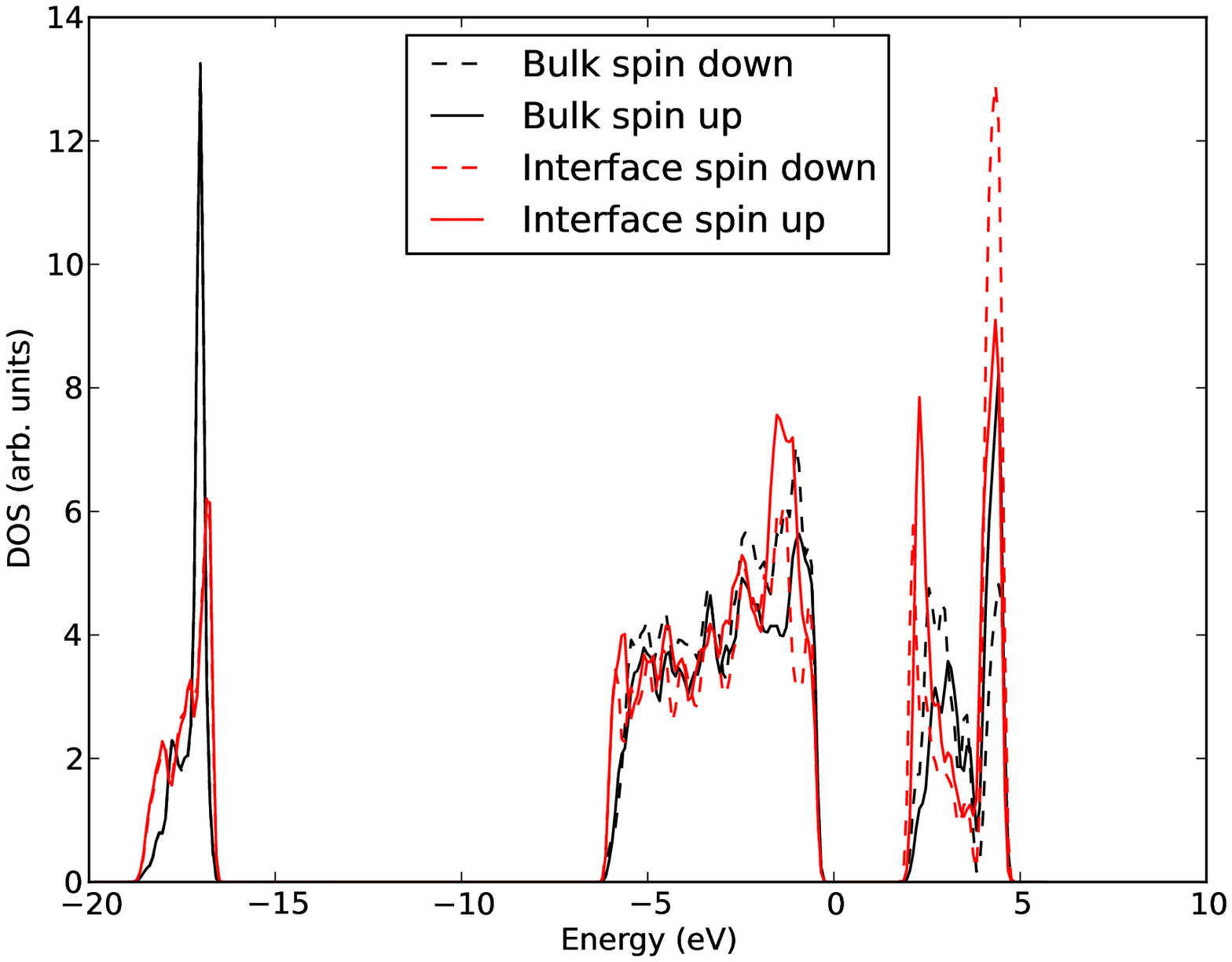}}
\subfigure[pyG(O)]{\label{img:DOSpyraG}\includegraphics[width=0.31\textwidth]{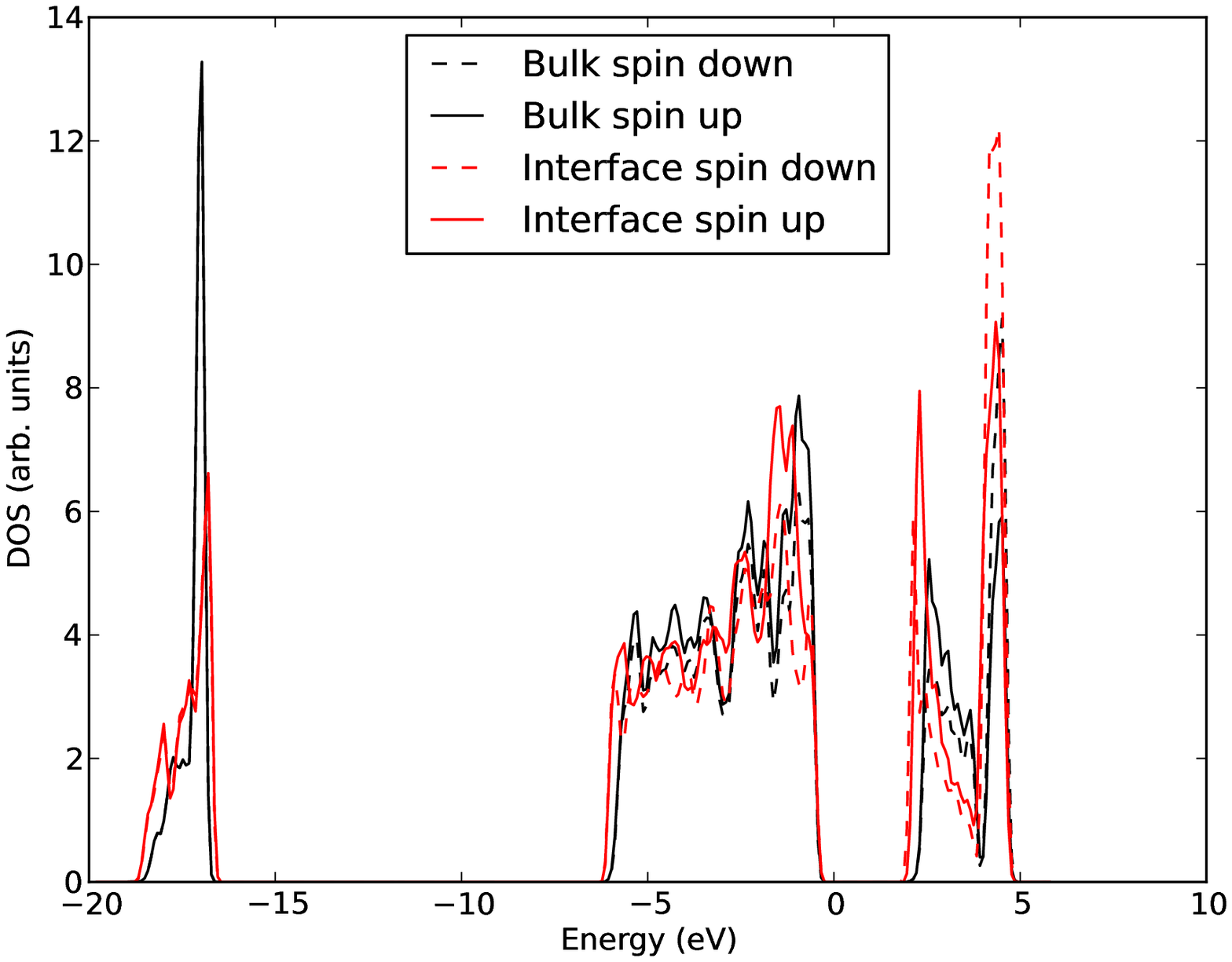}}
\caption{The density of states, DOS, of the GB interfaces.  The Fermi energy is at 0 eV.  The black curves correspond to the DOS of the bulk like regions of the interface, while the red curves are the DOS of the interface regions.  Solid lines represent spin up and dashed lines are spin down.   The GB represented here are the a) Rhombohedral Glide(V) and b) Rhombohedral Glide(O),  c) Rhombohedral Screw(V), d) Basal glide and e) Basal Glide-Mirror, f) Basal Rotational, g) Prismatic Screw, h) Prismatic Glide, i) Pyramidial Screw, and j) Pyramidial Glide interfaces. Color online. The projected DOS of the s, p, and d states are not shown to allow for comparison between the bulk and interface states.  However, the deep energy state at around 17.5 eV are dominated by O $s$ states and the states near the fermi energy are predominately O $p$ and Cr $d$ states (Cr $s$ states exist in this region as well, but the $d$ states dominate).}\label{img:DOS}
\end{figure*}

\subsection{Rhombohedral Grain Boundaries}

For the rhombohedral systems the energy ordering of the three symmetry/interfacial systems can be described based upon an analysis of the bond distances between the Cr-O bonds, Cr-Cr spacing, and O-O spacing in the first layers from the interface .  This is because all three of the systems considered for the rhombohedral GB (the O terminated interface with glide symmetry rG(O), the Cr vacancy terminated interface with glide symmetry rG(V), and the Cr vacancy terminated screw symmetry rS(V)) have anti-aligned spins at the interface. It causes the magnetic effects to only amplify the energy ordering resulting from the strained bond distances.  Starting with the least energetically favorable rhombohedral interface, rG(V), two features of the interfacial structure can be seen that affect the interfacial energy of the system.  First, the spacing between O atoms on either side of the interface is 2.24 $\AA$, which is $\sim$ 0.4 $\AA$ shorter than the smallest spacing seen in the bulk (2.66 $\AA$).  This is after an expansion of around 2\% in the [$\bar{5}052$]  direction during the optimization demonstrating the strain at this interface.  In addition there is a large number of strained Cr-O bonds (both compressed and stretched) at the interface with 2/3 of the bonds for each Cr atom at the interface strained by $\sim$0.1 $\AA$.   This is followed energetically by the rG(O) rhombohedral system, which contains reasonable O-O separation but has a large strain in the Cr-O bonds in the first couple of atomic layers of the interface.  Each Cr at the interface has 2/3 of its bonds strained by between 0.1 and 0.2 $\AA$, which corresponds to a 5-10\% strain in the bond.  Finally, the most energetically favorable rhombohedral system is the rS(V) system, which has a reasonable O - O bond and a lesser strain in the Cr-O interfacial bonds than the rG(O) system with a single bond (1/6) for each interfacial Cr stretched by 0.1 $\AA$.  This energy ordering and the corresponding analysis matches with Marinopoulos \emph{et al.} \cite{Marinopoulos2000} who observed the same energy ordering for Al$_2$O$_3$ due to O spacing.  They also see more strained bonding in the rG(V) and rG(O) systems compared to the rS(V).  This is shown within the density of states, DOS, in Figure \ref{img:DOSrglidev}-\ref{img:DOSrglideo} where the rG(V) and rG(O) systems both get significant changes in shape of the deep $s$ electrons and create spin polarized states composed of the $p$ and $d$ orbitals directly below the fermi energy. There is no major change of the DOS of rS(V) (Figure \ref{img:DOSrscrewv}) main over the same energies), which results in the low energy for the rS(V) GB.

\subsection{Basal Grain Boundaries}

The basal GB plane is particularly interesting because of the close structural relationship between the mirror, bM(O), and glide-mirror, bG(O), systems.  It is important to note that the bG(O) system during unconstrained optimization shifted to the bM(O) symmetry.  However, we felt that the bulk portion of each grain would prevent this shift in physical systems.  Therefore,  several layers in the "bulk" of each grain were constrained to the locations of the idealized symmetrical grain boundary.  This resulted in a gradual shift in atomic position between the two bulk regions with the atoms at the interface resembling the bM(O) interface as shown in Figure \ref{img:basal}b-c.  Since the bG(O) and the bM(O) interfaces are similar, it is the constraint of fixing the bulk regions of the bG(O) that shifts this system to slightly higher energies.  The difference in energy between the bM(O) interface and the bR(Cr) interface can be explained by the same structuring as seen by the Al$_2$O$_3$ basal interface~\cite{Marinopoulos2001}.  The bulk  corundum structure as shown in Figure \ref{img:planes} has a stacking of cations parallel to the [0001] direction of the form void-Cr($\uparrow$)-Cr($\downarrow$)-void (two vertical sets offset by a single step) with a spacing of 2.72 $\AA$ between Cr sites. However, for the basal GBs this stacking is changed.  In the bR(Cr) interface shown in Figure \ref{img:basal}a the two stacks change to one stack being void-Cr($\uparrow$)-void with the other becoming void-Cr($\uparrow$)-Cr($\downarrow$)-Cr($\downarrow$)-void with spacing between Cr of 2.58 $\AA$ and 2.63 $\AA$ (between similar spins).  For the bM(O) interface shown in Figure \ref{img:basal}b the first stack stays the same i.e. void-Cr($\uparrow$)-Cr($\downarrow$)-void with Cr spacing of 2.75 $\AA$, while the second stack becomes void-Cr($\uparrow$)-Cr($\downarrow$)-Cr($\uparrow$)-Cr($\downarrow$)-void with Cr spacing of 2.63 $\AA$, 2.46 $\AA$, and 2.63 $\AA$.   When the difference between the Cr spacing in each stack is compared we see that the bond distances are similar except for the addition of the 2.46 $\AA$ for the bM(O) interface.  This addition compression of 0.15 $\AA$ (effectively doubling the other distances) is expected to be the cause of the increase in E$_{int}$ for bM(O) in comparison to the bR(Cr) even with the two same spin neighbors in the bR(Cr) system.  The DOS of the basal plane GBs for Figure \ref{img:DOSbglide} - \ref{img:DOSbrot} supports this analysis with the bands beneath the fermi energies of the bM(O) and bG(O) systems shifting position by around an eV while the bR(Cr) system shifts band shape with a slight increase in energy with the band just below the fermi energy being spin polarized.   This large shift in the band positions  with respect to the bR(Cr) corresponds to the higher energy of the bM(O) and bG(O) systems.

\subsection{Prismatic Grain Boundaries}

Both prismatic GB planes are terminated in Cr-O planes.  For the glide plane interface, prG, the interface plane is in-between two Cr-O planes parallel to the [0001] direction, while for the screw plane interface, prS, the interface is at a Cr-O plane.  The biggest difference between the bulk and either the prG or the prS interfaces is the distribution of the Cr atoms within the O anion lattice.  In the bulk the Cr cations are distributed in a single plane in the (1010) direction in pairs of the form void-Cr($\uparrow$)-Cr($\downarrow$)-void with each plane offset by one place (i.e. the next plane has the distribution of Cr($\uparrow$)-Cr($\downarrow$)-void-Cr($\uparrow$)).  The intra-Cr spacing distance for the bulk is 2.959 $\AA$ with the two Cr offset slightly in the (0001) direction.   For  both the prG and the prS GBs, aside from minor changes in the O positions due to geometry optimization, the interface occurs where this distribution of Cr atoms has changed as shown in Figure \ref{img:pris}.  In the prG system the ABC plane distribution of the Cr atoms at the interface has planes of A=(void-void), B=(void-Cr($\uparrow$)-Cr($\downarrow$)-void), and C=(void-Cr($\uparrow$)-Cr($\downarrow$)-Cr($\uparrow$)-Cr($\downarrow$)-void).  It is worth noting that the middle two Cr atoms in both the B and C planes are not shifted in the (0001) direction.  This also increases the number of neighboring in-plane Cr atoms from 1 (for bulk) to 2 and 3 for the C plane interface cations.  Both of these effects are expected to raise the energy of the interface.  The prS interface also has an ABC plane distribution.  However, for prS the values are A = (void-Cr($\uparrow$)-void), B = (void-Cr($\uparrow$)-Cr($\downarrow$)-Cr($\uparrow$)-void), and C = (void).  For the B plane the middle Cr atom has 2 in-plane Cr neighbors, which causes a smaller increase in the interface energy than for the prG interface.  Therefore, the prS interface is the most energetically preferred prismatic twin grain boundary.  This agrees with the Al$_2$O$_3$ prismatic GBs as described by Fabris \emph{et al.} \cite{Fabris2002}  The DOS in Figure \ref{img:DOSprisS} and \ref{img:DOSprisG} for these two systems show a large deviation from the bulk regions of each interface with new states generated above both the $s$ and $p$/$d$ clusters of states.  These are due to the additional Cr neighbors due to the modified stacking order.  

\subsection{Pyramidal Grain Boundaries}

Like the other GB planes the pyramidal GB planes were selected from those considered in the Al$_2$O$_3$ literature~\cite{Fabris2001}.  This resulted in two glide symmetry systems, one with Cr termination, pyG(Cr), and one with O termination, pyG(O).  However, when these two structures were constructed for Cr$_2$O$_3$ the pyG(O) system shifted the first atomic layers upon optimization to become the pyG(Cr) interface.  It can therefore be concluded that unlike the Al$_2$O$_3$ structure pyG(O) is not stable in Cr$_2$O$_3$.   During the optimization of the pyG(O) system a quasi-stable system arose that resembled the Al$_2$O$_3$ pyG(O) system. This Cr$_2$O$_3$ system has a spacing of 2.51 $\AA$ between interfacial Cr, compared to the 2.72 $\AA$ in bulk, and an inter-O spacing of 2.02 $\AA$ compared to the 2.66 $\AA$ minimum distance between O atoms in the bulk.  These highly strained inter-Cr and inter-O spacing is the cause of the pyG(O) interface shifting to the pyG(Cr) interface.  However, this is not strongly portrayed in the DOS as shown in Figure \ref{img:DOSpyraS} and \ref{img:DOSpyraG} where the DOS do not significantly change in relation to the bulk structures. 

\subsection{Energy Ordering of  Grain Boundaries}

Using \ref{eqn:eint} to calculate the interfacial energy allows for comparison among the interfaces.   Particularly, it allows for the comparison of the energies of the surface bonding and interface splitting that would be major determining factors in which interfaces exist at grain boundaries.  As can be seen in table \ref{tab:gbenergy2} the six most energetically favorable interfaces are the prS $<$ rS(V) $<$ prG $<$ bR(Cr) $<$ rG(O) $<$ pyG(Cr).  This is consistent with the calculations of Fang \emph{et al.} \cite{Fang2012}, which has the prismatic interface energy being at lower energy than either the Cr or O terminated basal interfaces.  
For the Cr$_2$O$_3$ GBs two of the first three most favorable systems are the prismatic symmetries considered because the O lattice is effectively unchanged for the prismatic systems and only the Cr ion stacking (in the [10$\bar{1}$0] direction) is changed as discussed above.  This effectively changes the bonding structure of the Cr without changing the O lattice. Conversely, the rhombohedral system, which is the second most favorable system, is characterized by moderate strain at the interface demonstrated by the single strained Cr-O bond for the interfacial Cr and the O, which does not significantly modify the DOS.  This results in an energetically stable structure between the two prismatic systems.  The energetically favorable basal plane system follows the prismatic and rhombohedral interfaces because the rotational symmetry system combines the strained interface with a Cr ordering change (this time in the [0001] direction).  Finally, the single stable pyramidal plane interface is of the highest energy of the four planes considered because of the high number of strained atomic planes as discussed above with both the Cr-Cr interfacial plane distance and the O-O interfacial planes distance compressed between 0.2 $\AA$ and 0.4 $\AA$.

\section{Conclusion}

In conclusion, we have studied twin grain boundaries with multiple symmetries for each of the rhombohedral, basal, prismatic, and pyramidal planes based upon the structures from the literature for Al$_2$O$_3$, which has the same corundum structure followed by a DOS analysis. From this we see that the prismatic screw with a Cr-O plane interface is energetically the most preferred system. The rhombohedral screw symmetry interface with vacancy termination, the prismatic glide symmetry interface and the basal rotational symmetry interface being the second, third and fourth energetically preferred systems respectively, with a range of $\sim$0.4 J/m$^2$  among them. The remaining interfaces then start with at least this much higher energy than the basal rotational system. This suggests that other systems are significantly less likely than these four preferred systems, which therefore means that these four interfaces will be the most common twin grain boundaries found in poly-crystalline Cr$_2$O$_3$. Further the DOS reveal that the prismatic systems have higher polarized defect states than the rhombohedral screw interface.

 Post-doctoral grants of the Electricite de France (EDF) for the support of A.G. Van Der Geest and M. M. Islam are acknowledged.

\section*{References}
\bibliographystyle{unsrt}

\begin{thebibliography}{10}

\bibitem{Diawara2010}
Boubakar Diawara, Yves-Alain Beh, and Philippe Marcus.
\newblock Nucleation and growth of oxide layers on stainless steels (fecr)
  using a virtual oxide layer model.
\newblock {\em The Journal of Physical Chemistry C}, 114(45):19299--19307,
  2010.

\bibitem{Yu2012}
Haobo Yu, Changfeng Chen, Ruijing Jiang, Ping Qiu, and Yujing Li.
\newblock Migration of ion vacancy in hydroxylated oxide film formed on cr: A
  density functional theory investigation.
\newblock {\em The Journal of Physical Chemistry C}, 116(48):25478--25485,
  2012.

\bibitem{Tsai1996}
S.C. Tsai, A.M. Huntz, and C.~Dolin.
\newblock Growth mechanism of cr2o3 scales: oxygen and chromium diffusion,
  oxidation kinetics and effect of yttrium.
\newblock {\em Materials Science and Engineering: A}, 212(1):6 -- 13, 1996.

\bibitem{Fang2012}
H~Z Fang, Y~Wang, S~L Shang, Paul~D Jablonski, and Z~K Liu.
\newblock First-principles calculations of interfacial and segregation energies
  in α-cr 2 o 3.
\newblock {\em Journal of Physics: Condensed Matter}, 24(22):225001, 2012.

\bibitem{Catlow1989}
CRA Catlow, SC~Parker, and MP~Allen.
\newblock {\em Computer Modelling of Fluids, Polymers and Solids}.
\newblock Dordrecht: Kluwer Academic, 1989.

\bibitem{Marinopoulos2000}
A.G Marinopoulos and C~Els{\"a}sser.
\newblock Microscopic structure and bonding at the rhombohedral twin interface
  in α-al2o3.
\newblock {\em Acta Materialia}, 48(18--19):4375 -- 4386, 2000.

\bibitem{Marinopoulos2001}
A.~G. Marinopoulos, S.~Nufer, and C.~Els\"asser.
\newblock Interfacial structures and energetics of basal twins in
  $\alpha$-al$_{2}$o$_{3}$: First-principles density-functional and empirical
  calculations.
\newblock {\em Phys. Rev. B}, 63:165112, Apr 2001.

\bibitem{Fabris2002}
Stefano Fabris, Stefan Nufer, Christian Els\"asser, and Thomas Gemming.
\newblock Prismatic $\sigma$3(10$\bar{1}$0) twin boundary in
  $\alpha$-al$_{2}$o$_{3}$ investigated by density functional theory and
  transmission electron microscopy.
\newblock {\em Phys. Rev. B}, 66:155415, Oct 2002.

\bibitem{Fabris2001}
Stefano Fabris and Christian Els\"asser.
\newblock $\sigma$13(10$\bar{1}$4) twin in $\alpha$-al$_{2}$o$_{3}$: A model
  for a general grain boundary.
\newblock {\em Phys. Rev. B}, 64:245117, Dec 2001.

\bibitem{Kresse1996b}
G.~Kresse and J.~Furthm{\"u}ller.
\newblock Efficiency of ab-initio total energy calculations for metals and
  semiconductors using a plane-wave basis set.
\newblock {\em Computational Materials Science}, 6(1):15 -- 50, 1996.

\bibitem{Kresse1996}
G.~Kresse and J.~Furthm\"uller.
\newblock Efficient iterative schemes for \textit{ab initio} total-energy
  calculations using a plane-wave basis set.
\newblock {\em Phys. Rev. B}, 54:11169--11186, Oct 1996.

\bibitem{perdew1993}
John~P. Perdew, J.~A. Chevary, S.~H. Vosko, Koblar~A. Jackson, Mark~R.
  Pederson, D.~J. Singh, and Carlos Fiolhais.
\newblock Erratum: Atoms, molecules, solids, and surfaces: Applications of the
  generalized gradient approximation for exchange and correlation.
\newblock {\em Phys. Rev. B}, 48:4978--4978, Aug 1993.

\bibitem{Blochl1994}
P.~E. Bl\"ochl.
\newblock Projector augmented-wave method.
\newblock {\em Phys. Rev. B}, 50:17953--17979, Dec 1994.

\bibitem{Monkhorst1976}
Hendrik~J. Monkhorst and James~D. Pack.
\newblock Special points for brillouin-zone integrations.
\newblock {\em Phys. Rev. B}, 13(12):5188--5192, Jun 1976.

\bibitem{dudarev1998}
S.~L. Dudarev, G.~A. Botton, S.~Y. Savrasov, C.~J. Humphreys, and A.~P. Sutton.
\newblock Electron-energy-loss spectra and the structural stability of nickel
  oxide:$\hskip0.3em${}$\hskip0.3em${}an lsda+u study.
\newblock {\em Phys. Rev. B}, 57:1505--1509, Jan 1998.

\bibitem{Rohrbach2004}
A.~Rohrbach, J.~Hafner, and G.~Kresse.
\newblock Ab initio study of the (0001) surfaces of hematite and chromia:
  Influence of strong electronic correlations.
\newblock {\em Phys. Rev. B}, 70:125426, Sep 2004.

\end{thebibliography}

\end{document}